\def\ubm{$U_{\rm m}-B_{\rm m}$}
\def\bvm{$B_{\rm m}-V_{\rm m}$}
\def\dt{$\Delta{\rm log}\,T$}
\def\dl{$\Delta{\rm log}\,L$}
\def\av{$A_{V}$}

\documentclass[aaspp4,11pt,preprint,apjfonts]{emulateapj}

\shorttitle{The SED of Post-Starburst Galaxies in the NMBS}
\shortauthors{Kriek et al.}

\begin{document}
  
\title{The Spectral Energy Distribution of Post-Starburst Galaxies in the NEWFIRM Medium-Band Survey: A Low Contribution from TP-AGB Stars}

\author{Mariska Kriek\altaffilmark{1}, Ivo Labb\'e\altaffilmark{2}, Charlie Conroy\altaffilmark{1,3},  Katherine E. Whitaker\altaffilmark{4}, Pieter G. van Dokkum\altaffilmark{4}, Gabriel B. Brammer\altaffilmark{4}, Marijn Franx\altaffilmark{5}, Garth D. Illingworth\altaffilmark{6}, Danilo Marchesini\altaffilmark{7}, Adam Muzzin\altaffilmark{4}, Ryan F. Quadri\altaffilmark{5}, \& Gregory Rudnick\altaffilmark{8}}

\altaffiltext{1}{Department of Astrophysical Sciences, Princeton University, 
 Princeton, NJ 08544, USA}

\altaffiltext{2}{Carnegie Observatories, 813 Santa Barbara Street, Pasadena, CA 91101, USA}

\altaffiltext{3}{Harvard-Smithsonian Center for Astrophysics, 60 Garden Street, Cambridge, MA 02138, USA}

\altaffiltext{4}{Department of Astronomy, Yale University, New Haven, 
  CT 06520, USA}

\altaffiltext{5}{Leiden Observatory, Leiden University, NL-2300 RA Leiden, 
  The Netherlands}

\altaffiltext{6}{UCO/Lick Observatory, University of California, Santa Cruz, CA 95064, USA}

\altaffiltext{7}{Department of Physics and Astronomy, Tufts University, Medford, MA 02155, USA}

\altaffiltext{8}{Department of Physics and Astronomy, The University of Kansas, Lawrence, KS 66045, USA}

\begin{abstract}
Stellar population synthesis (SPS) models are a key ingredient of many
galaxy evolution studies. Unfortunately, the models are still poorly
calibrated for certain stellar evolution stages. Of particular concern
is the treatment of the thermally-pulsing asymptotic giant branch
(TP-AGB) phase, as different implementations lead to systematic
differences in derived galaxy properties. Post-starburst galaxies are
a promising calibration sample, as TP-AGB stars are thought to be most
prominently visible during this phase. Here, we use post-starburst
galaxies in the NEWFIRM medium-band survey to assess different SPS
models. The available photometry allows the selection of a homogeneous
and well-defined sample of 62 post-starburst galaxies at
$0.7\lesssim\,z\lesssim\,2.0$, from which we construct a well-sampled
composite spectral energy distribution (SED) over the range
$1200-40\,000$~\AA. The SED is well fit by the Bruzual \& Charlot SPS
models, while the Maraston models do not reproduce the rest-frame
optical and near-infrared parts of the SED simultaneously. When the
fitting is restricted to $\lambda<6000$~\AA, the Maraston models
overpredict the near-infrared luminosity, implying that these models
give too much weight to TP-AGB stars. Using the flexible SPS models by
Conroy et al, and assuming solar metallicity, we find that the
contribution of TP-AGB stars to the integrated SED is a factor of
$\sim3$ lower than predicted by the latest Padova TP-AGB
models. Whether this is due to lower bolometric luminosities, shorter
lifetimes, and/or heavy dust obscuration of TP-AGB stars remains to be
addressed. Altogether, our data demand a low contribution from TP-AGB
stars to the SED of post-starburst galaxies.
\end{abstract} 


\section{INTRODUCTION}\label{sec:int}

Galaxy evolution studies at all epochs strongly rely on stellar
population synthesis (SPS) models
\citep[e.g.,][]{tg76,le99,bc03,ma05,co09}. These models are key in
deriving  stellar masses, stellar population properties, and in some
cases even redshifts from spectra and broadband photometry. Thus, our
current understanding of stellar populations and galaxy growth across
cosmic time \citep[e.g.,][]{pa01,fo04,sh05,kr09b,vd10}, and the
evolution of the stellar mass density
\citep[e.g.,][]{ru06,st07,ma09,go10,la10} is critically dependent on
SPS models.

Nonetheless, SPS models are still poorly calibrated for certain
stellar evolution stages. In particular, the thermally pulsing
asymptotic giant branch (TP-AGB) phase is a source of major
discrepancy among different SPS models; different implementations of
its treatment lead to large systematic differences in derived galaxy
properties. TP-AGB stars are cool giants ($T\lesssim 4000$\,K) and
primarily contribute to the near-infrared part of the spectrum. They
are of low-to-intermediate stellar mass, and thus their relative
contribution to the near-infrared spectrum is highest around
0.5--2.0~Gyr after a starburst. Due to the short lifetime of AGB stars
($\sim$1\,Myr), the local globular cluster calibration samples are
small, and the total luminosity, temperature, and carbon star fraction
in the TP-AGB phase are not well calibrated. For example, the TP-AGB
phase is more prominent in the \cite{ma05} models than in
\cite{bc03}. Therefore, stellar masses derived using the \cite{bc03}
models are generally higher than stellar masses derived using the
\cite{ma05} SPS models \citep[e.g.,][]{wu07,mu09,kg07,we06}.

Several studies use galaxy spectral energy distributions (SEDs) to
assess the different SPS models, and resolve the
discrepancies \citep[e.g,][]{ma06,em08}. This exercise is complicated
due to the large variety of possible stellar populations and the
degeneracies between the star formation timescale, age, metallicity,
and dust. However, specific galaxy populations can be used to
constrain certain stellar evolution phases. For example,
\cite{cg10} use post-starburst galaxies in the Sloan Digital Sky
Survey (SDSS) to assess the treatment of the TP-AGB phase of different
SPS models, as TP-AGB stars dominate the near-infrared luminosity
during this phase. They find that the models by \cite{bc03} and
\cite{co09} can reproduce the optical and near-infrared colors of
post-starburst galaxies, while the \cite{ma05} models cannot.

In this Letter, we take a similar approach as \cite{cg10} and use a
post-starburst galaxy population in the NEWFIRM medium-band
survey \citep[NMBS;][]{vd09a} to test different SPS models, and obtain
new constraints on the SED shape during the time that the TP-AGB stars
are thought to be most dominant. The NMBS allows us to photometrically
select a large and clean sample of post-starburst galaxies, and to
construct a composite SED which is more thoroughly sampled than in
\cite{cg10}. Throughout the Letter, we assume a $\Lambda$CDM cosmology
with $\Omega_{\rm m}=0.3$, $\Omega_{\rm\Lambda}=0.7$, and $H_{\rm
  0}=70$~km~s$^{-1}$ Mpc$^{-1}$.

\section{SAMPLE SELECTION}

\begin{figure}  
  \begin{center} 
    \includegraphics[width=0.48\textwidth]{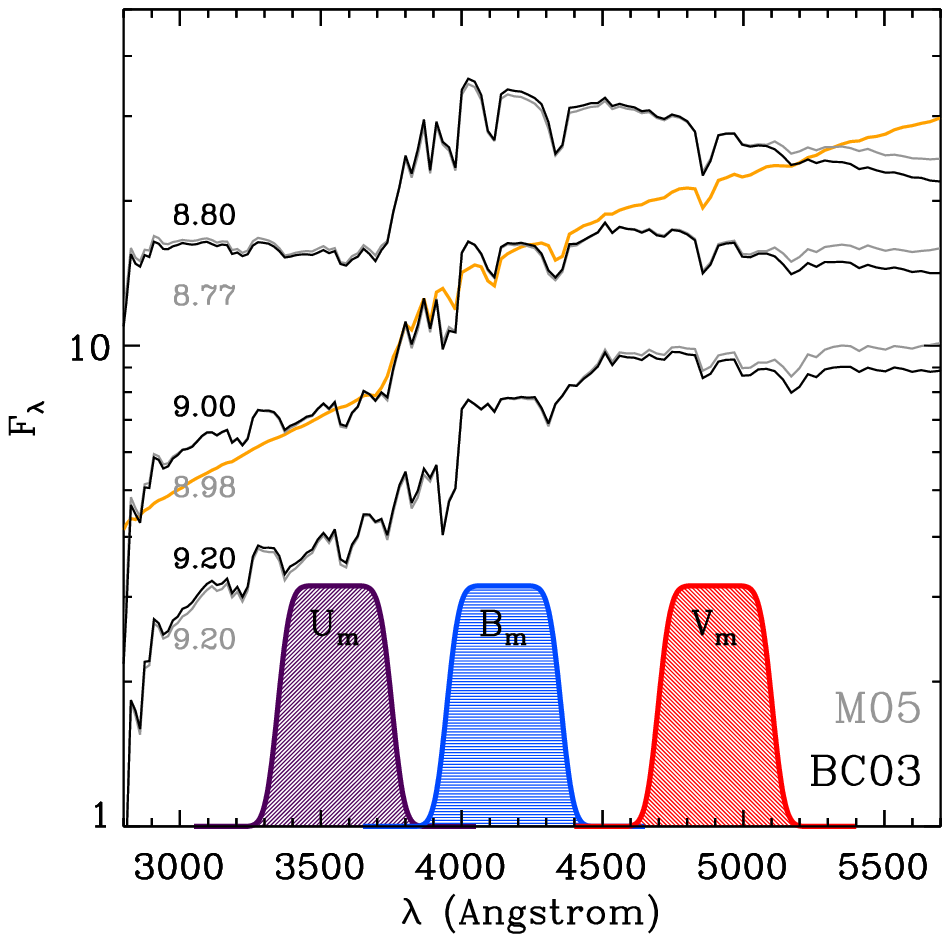} 
    
    \caption{Synthetic filter curves of intermediate bandwidth used
      to select post-starburst galaxies. \ubm\ isolates the Balmer
      break, while \bvm\ measures the slope redwards of the break. The
      \cite{bc03} and \cite{ma05} models for the same IMF \citep{sa55}
      are shown for ages of log\,($t$/yr)\,$\approx$\,8.8,\,9.0, and
      9.2, and an e-folding time of log\,($\tau$/yr)\,=\,8.0. We also
      show an SED of a dusty star-forming galaxy ($\tau=1$\,Gyr and
      $t=100$\,Myr) with the same \ubm\ color as the 1~Gyr old
      post-starburst galaxy (orange). All three filters fall in the
      regime where the model colors of post-starburst galaxies are
      nearly identical, and thus our selection is not biased toward
      any particular SPS model. \label{fig:ill}}
    
  \end{center}          
\end{figure} 

\begin{figure}   
  \begin{center} 
    \includegraphics[width=0.48\textwidth]{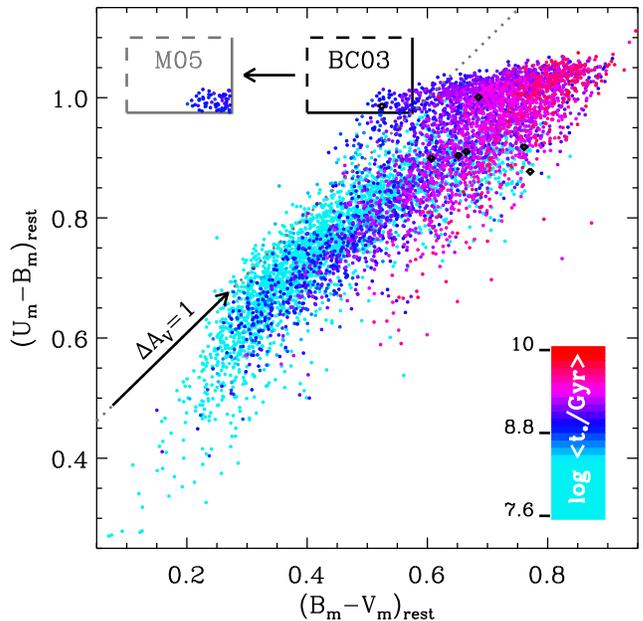}
    
    \caption{Rest-frame \ubm\ vs. \bvm\ for all $z>0.5$ galaxies in the
      NMBS (with (S/N)$_{K}>25$). The black box indicates our
      post-starburst selection, characterized by red \ubm\ colors and
      relatively blue \bvm\ colors. The dust vector and dotted line
      indicate that contamination by obscured star-forming galaxies is
      minimal. The galaxies are color coded following their
      \cite{bc03} average stellar ages. In the gray box, we show the
      \cite{ma05} average stellar ages of the selected galaxies. For
      both models the rest-frame near-infrared is down-weighted when
      fitting. The black diamonds are the seven galaxies in
      \cite{ma06}.\label{fig:sel}} 
    
  \end{center}           
\end{figure}  

\begin{figure*}   
  \begin{center}     
    \includegraphics[width=0.9\textwidth]{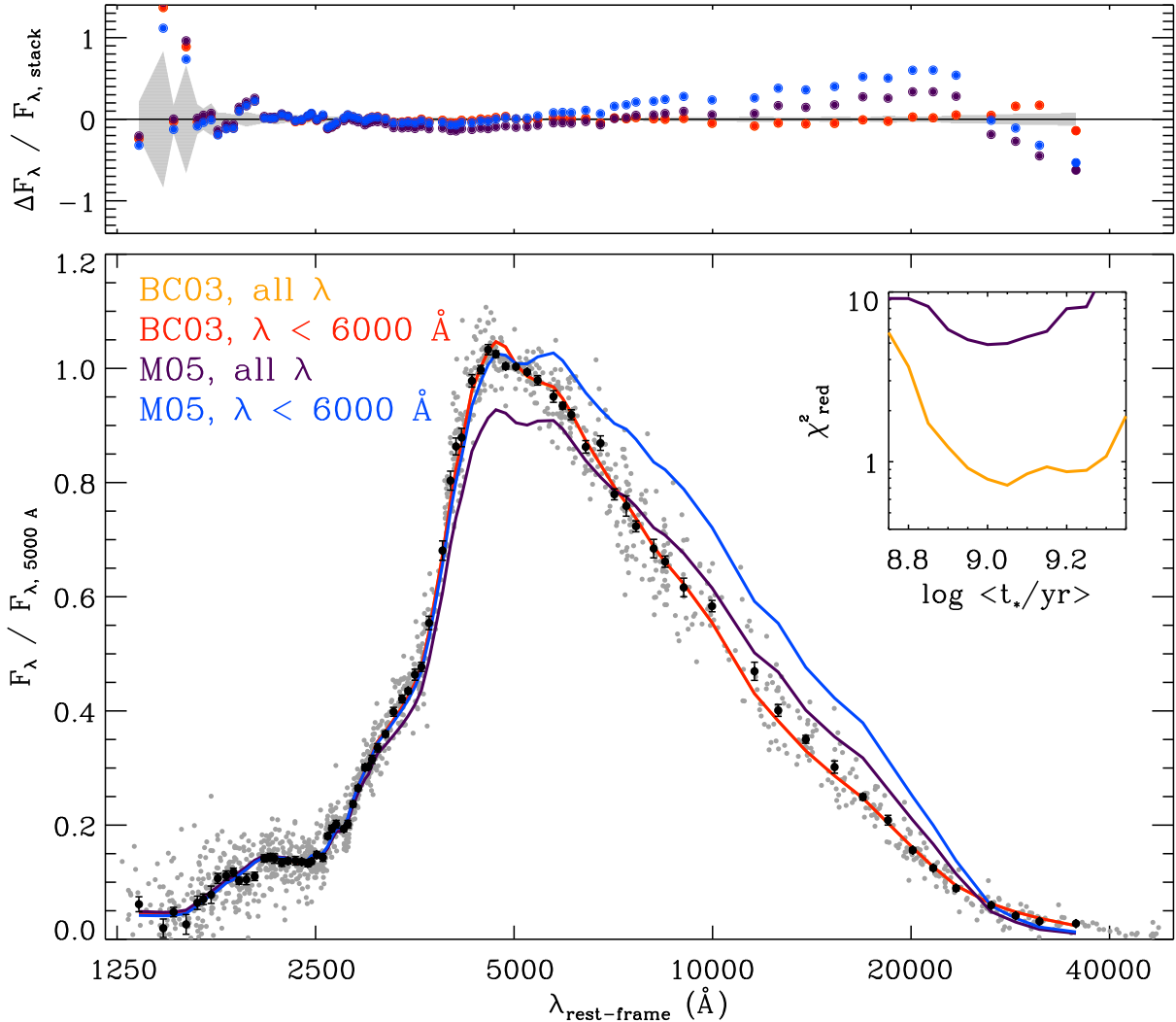}
    
    \caption{Composite post-starburst SED (black dots), derived
      from the individual SEDs (gray dots) of 62 post-starburst
      galaxies in the NMBS. The individual SEDs are normalized at
      5000~\AA\ before combining. The best-fit \cite{bc03} and
      \cite{ma05} SPS models to the full spectrum, with the same
      resolution as the data, are shown by the orange and purple
      curves, respectively. The red and blue curves show the
      respective best fits when excluding $\lambda>6000$~\AA\ in the
      fit. The inset represents the minimum $\chi^2$ value per degree
      of freedom vs. the mean stellar age for both SPS models. In the
      top panel, we show the fractional offset between the models and
      the data. The gray shaded area indicates the uncertainties on
      the composite SED. The \cite{bc03} models more accurately
      reproduce the SED shape of post-starburst galaxies than the
      \cite{ma05} models. \label{fig:sed}}
    
  \end{center}           
\end{figure*}

The NMBS uses five custom near-infrared medium-band filters and covers
a total area of 0.4 deg$^2$ in the COSMOS \citep{sc07} and
AEGIS \citep{da07} fields. The medium-band filter set, in combination
with deep optical medium and broadband photometry and IRAC imaging,
provides accurate photometric redshifts and stellar population
properties \citep[e.g.,][]{br09,vd10,wh10}. A full overview of this
survey is given in K. E. Whitaker et al. (2010, in preparation). The
photometric redshifts and stellar population properties are derived
using EAZY \citep{br08} and FAST \citep{kr09a}, respectively.

Our post-starburst galaxy sample is selected using three optimized
synthetic rest-frame filters of intermediate
bandwidth\footnote{Rest-frame colors are computed from the best-fit
EAZY templates \citep[see][]{br09}.} (see Figure~\ref{fig:ill}). The
$U_{\rm m}$ and $B_{\rm m}$ filters isolate the Balmer break,
while \bvm\ measures the slope of the SED just redwards of the Balmer
break. The location of the $V_{\rm m}$ filter is a trade-off between
being red enough to constrain the SED slope and blue enough not to
enter the regime where the stellar evolution models and spectral
libraries start to deviate \citep[see also][]{m09}. Post-starburst
galaxies have strong Balmer breaks, characterized by red \ubm\ colors
and blue \bvm\ colors, and thus are expected to lie in the black
selection box in Figure~\ref{fig:sel}. The boundaries of the selection
box are a trade-off between minimizing contamination by dusty
star-forming galaxies (estimated to be $\sim$1\%) and obtaining a
large enough sample. As indicated by the orange SED in
Figure~\ref{fig:ill}, dusty star-forming galaxies with the same \ubm\
color as post-starburst galaxies will have redder \bvm\ colors.

In order to solely include high-quality SEDs with rest-frame UV-to-NIR
wavelength coverage, we require a redshift of $z>0.5$ and a
$K$-band signal-to-noise ratio (S/N) of 25. We also exclude galaxies
for which the IRAC fluxes are strongly contaminated by nearby sources,
and thus cannot properly be deblended \citep{la05}. The final
selection consists of 62 galaxies at $0.695<z<2.028$, with a median
redshift of $z=1.56$. Six targets have spectroscopic redshifts,
yielding a median photometric redshift uncertainty of $\Delta
z/(1+z)=0.005$.

We have inspected the SEDs of all targets by eye and none have obvious
problems. Furthermore, for none of the selected galaxies there are
indications that the rest-frame UV-to-NIR continuum is dominated by
dusty star formation (Figure~\ref{fig:sel}). Similar to
spectroscopic post-starburst galaxy samples, obscured star formation
may have been missed \citep[e.g.,][]{mo01,go07}. However, some
contamination by a dusty star-forming population will not compromise
our results (see Section 4.1).

To obtain a homogeneous sample and to avoid modeling degeneracies, our
selection intentionally excludes younger or heavily dust-obscured
post-starburst galaxies and specifically targets galaxies with short
star formation timescales, for which the TP-AGB phase is most
prominent. Therefore, our sample is likely more restrictive than other
spectroscopically selected post-starburst galaxy samples
\citep[e.g.,][]{qu04}. However, for the purpose of this Letter it is
not relevant whether our sample is complete, as long as the selection
is not biased toward any particular SPS model.

\begin{deluxetable*}{l l l c c c c c c c c} 
  \tabletypesize{\scriptsize}  \tablecaption{Stellar Population
    Properties \label{tab:mod}} \tablewidth{0pt}   \tablehead{ & & &
    \colhead{log\,$t$} & \colhead{log\,$\tau$} & \colhead{\av} & &
    \colhead{log\,$M_*$} & \colhead{log SFR} & \colhead{log SFR/$M_*$} &  
      \\ \colhead{ID}  & \colhead{Fit
      region} & \colhead{Model} & \colhead{(yr)} & \colhead{(yr)} &
    \colhead{(mag)} & \colhead{$Z$} & \colhead{$M_{\odot}$} &
    \colhead{$M_{\odot}$\,yr$^{-1}$} & \colhead{yr$^{-1}$} & 
    \colhead{$\chi^2_{\rm red}$}}    
  \startdata  

  Composite SED &            Full range &  BC03 &    9.06$^{+
    0.01}_{-   0.01 }$ &    8.10$^{+    0.04}_{-    0.02 }$ &
  0.16$^{+    0.02}_{-    0.07 }$ &   0.020 & 11.10$^{+    0.00}_{-
    0.02 }$ &   -0.63$^{+    0.01}_{-    0.10 }$ &  -11.73$^{+
    0.00} _{-    0.08 }$ &    0.73\\ &                       &   M05 &
  9.05$^{+    0.00}_{-   0.00 }$ &    8.10$^{+    0.00}_{-    0.00 }$
  &    0.00$^{+    0.00}_{-    0.00 }$ &   0.020 & 10.99$^{+
    0.00}_{-    0.00 }$ &   -0.67$^{+    0.00}_{-    0.00 }$ &
  -11.66$^{+    0.00} _{-    0.00 }$ &    5.08\\ & $\lambda <$
  6000\,\AA &  BC03 &    9.06$^{+    0.18}_{-   0.02 }$ &    8.10$^{+
    0.20}_{-    0.05 }$ &    0.16$^{+    0.15}_{-    0.12 }$ &   0.020
  & 11.10$^{+    0.09}_{-    0.05 }$ &   -0.63$^{+    0.18}_{-    0.13
  }$ &  -11.73$^{+    0.14} _{-    0.08 }$ &    0.72\\ &
  &   M05 &    9.06$^{+    0.01}_{-   0.01 }$ &    8.10$^{+
    0.08}_{-    0.03 }$ &    0.08$^{+    0.08}_{-    0.08 }$ &   0.020
  & 11.09$^{+    0.03}_{-    0.12 }$ &   -0.65$^{+    0.10}_{-    0.12
  }$ &  -11.74$^{+    0.20} _{-    0.08 }$ &    0.94\\ Ma-3650 &
  Full range &  BC03 &    9.08$^{+    0.07}_{-   0.19 }$ &    8.10$^{+
    0.14}_{-    1.10 }$ &    0.40$^{+    0.45}_{-    0.13 }$ &   0.008
  & 10.89$^{+    0.13}_{-    0.00 }$ &   -1.03$^{+    0.47}_{-   27.34
  }$ &  -11.92$^{+    0.42} _{-   27.45 }$ &    2.91\\ $z_{\rm
    spec}=1.91$ &                       &   M05 &    8.94$^{+
    0.11}_{-   0.17 }$ &    7.90$^{+    0.22}_{-    0.90 }$ &
  0.10$^{+    0.13}_{-    0.10 }$ &   0.020 & 10.76$^{+    0.03}_{-
    0.11 }$ &   -1.54$^{+    0.73}_{-   20.98 }$ &  -12.30$^{+
    0.70} _{-   20.97 }$ &    2.90\\

  \enddata

  \tablecomments{We fit a grid with log\,($t$/yr) between 7.6 and
    10.1, in steps of 0.01 (not exceeding the age of the universe),
    log\,($\tau$/yr) between 7 and 10 in steps of 0.1, and
    \av\ between 0 and 3 in steps of 0.02\,mag. For each model we
    explore three different metallicities: subsolar (0.01 and
    0.008 for the \cite{ma05} and \cite{bc03} models, respectively),
    solar, or supersolar (0.04 and 0.05 for the \cite{ma05} and
    \cite{bc03} models, respectively). To account for systematic
    errors, we increase all photometric uncertainties by quadratically
    adding 5\% of the photometric flux. The errors represent the 68\%
    confidence intervals.}  

\end{deluxetable*}

\section{COMPOSITE SPECTRUM}

In order to make a composite spectrum, we deredshift the SEDs of all
galaxies to rest frame, without changing the observed fluxes. All SEDs
are normalized at 5000~\AA, the longest wavelength for which the
variations between the different SPS models are negligible (see
Figure \ref{fig:ill}). We create a composite SED by averaging the
rest-frame fluxes in wavelength bins of 20 data points each. The
uncertainties on the composite spectrum are determined by
bootstrapping.

Figure~\ref{fig:sed} shows the composite post-starburst galaxy SED in
black. The photometric data points of the individual galaxies are
represented by the gray dots. The uniformity of all individual SEDs is
remarkable and results in a high-quality, low-resolution spectrum of
a post-starburst galaxy. Even subtle features, such as the Mg{\sc\,ii}
absorption line at 2800~\AA\ and the continuum break at 2640~\AA, are
detected.

\section{COMPARISON TO MODELS}\label{sec:mod}

\subsection{\cite{bc03} and \cite{ma05}}

The high quality of the composite post-starburst galaxy SED allows the
assessment of the different SPS models. We fit the composite SED by
both the \cite{bc03} and \cite{ma05} models, assuming a star formation
history (SFH) of the form $\psi(t)\propto{\rm exp}(-t/\tau)$ and
leaving age ($t$), the $e$-folding time ($\tau$), the amount of dust
attenuation (\av), and metallicity ($Z$) as free parameters (see note
to Table~\ref{tab:mod}). We assume the \cite{ca00} attenuation curve,
which we implement as a uniform screen. We adopt the \cite{sa55}
initial mass function (IMF), as this IMF is available for both SPS
models. Compared to a \cite{kr01} or \cite{ch03} IMF our choice will
primarily affect the mass-to-light ratio and has little impact on
other stellar population properties. For each bin, we determine the
effective filter curve, by adding the deredshifted filter curves of
all included data points. In most cases, this is a variety of
broadband and medium-band filters with different observed wavelengths,
as our sample spans a large range in redshift.

We start by fitting the full wavelength range. The best fits for
\cite{bc03} and \cite{ma05} are represented by the orange and purple
fits in Figure~\ref{fig:sed}, respectively.  The \cite{bc03} models
yield an acceptable fit to the data, while the \cite{ma05} models do
not fit the full wavelength range simultaneously. The best-fit stellar
population properties are broadly consistent, with a slightly lower
value for stellar mass ($M_{*}$) and \av\ for the \cite{ma05} models
(Table~\ref{tab:mod}). We have explored other SFHs as well, such as a
delayed exponential SFH ($\psi(t)\sim t~{\rm exp}(-t/\tau)$), a
truncated SFH preceded by a constant star formation rate (SFR, with
$\tau=t_{\rm stop}-t_{\rm start}$), a truncated SFH preceded by an
exponentially increasing SFR, and a two-component population, but none
provide a significantly better correspondence between the \cite{ma05}
models and the composite SED. The SFH used to generate the model SEDs
in Figure~\ref{fig:sed} produces a sharply peaked spectrum; adding an
older or dusty star-forming population would only broaden the model
SED, which would then produce an even larger discrepancy between the
\cite{ma05} models and the data.

To better understand the discrepancies between the models in the
rest-frame near-infrared, we repeat the fitting, but now restricting
the wavelength range to $\lambda<6000$~\AA. The \cite{bc03} and
\cite{ma05} models, as represented by the red and blue curves,
respectively, fit this wavelength range nearly equally well and yield
consistent values for $M_*$, $t$, $\tau$, and
\av\ (Table~\ref{tab:mod}). However, while \cite{bc03} also fit the
full wavelength coverage, the \cite{ma05} models overpredict the
rest-frame near-infrared flux. 

\begin{figure*} 
  \begin{center} 
    \includegraphics[width=0.9\textwidth]{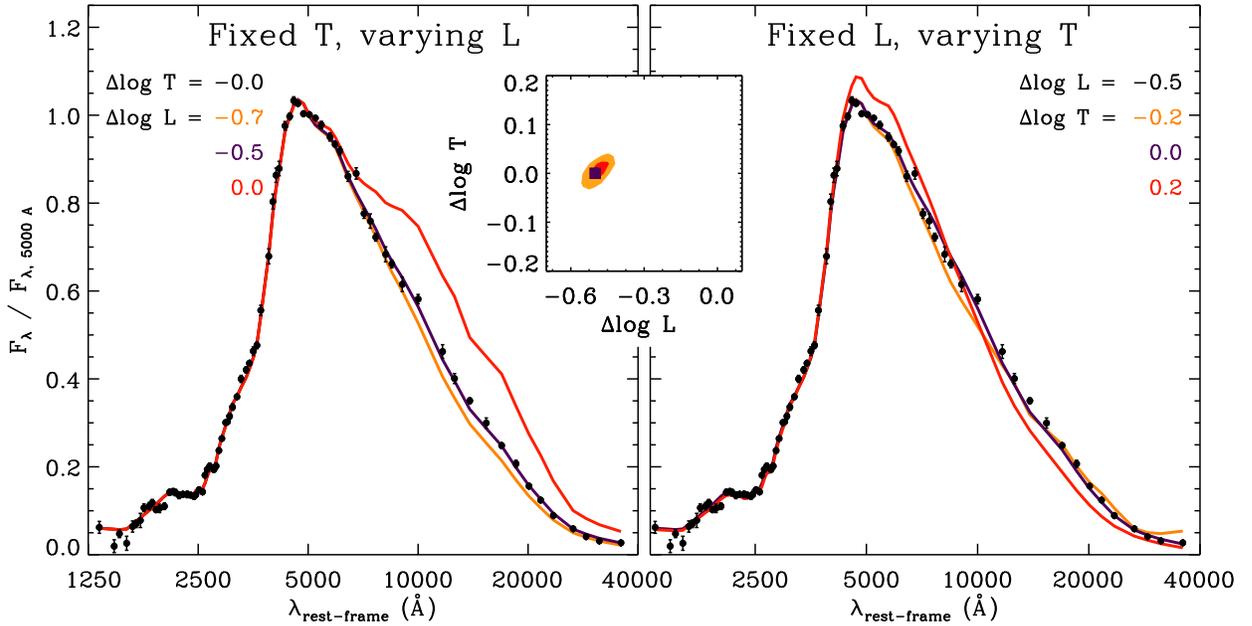}

    \caption{Comparison of the composite post-starburst galaxy
      spectrum with the FSPS models by \cite{co09}, for which the
      temperature and bolometric luminosity of the TP-AGB stars can be
      modified with respect to the Padova stellar evolution models
      (\dt\ and \dl, respectively). The left panel illustrates the
      influence of varying the luminosity, while keeping the
      temperature fixed to the best-fit value. In the right panel, the
      luminosity is fixed to the best-fit value, but the temperature
      varies. In both cases, the best fit is shown in purple. For both
      panels, $t$, $\tau$, and \av\ are fixed as well. The inset shows
      the best-fit \dt\ and \dl\ (purple square), and its
      corresponding 68\% and 95\% confidence
      intervals. \label{fig:fsps}}

  \end{center} 
\end{figure*}

\subsection{Tests and Comparison to other Studies}

As a check on our procedure, we stack the best fits to the individual
SEDs for both models. The stacks are similar to the purple and orange
curves in Figure~\ref{fig:sed}. Along the same lines, we assess the
qualities of the individual fits. All galaxies favor the
\cite{bc03} models, with median values for $\chi^2_{\rm BC03}$ and
$\chi^2_{\rm M05}$ of 0.91 and 4.3, respectively.  

Whereas both the \cite{bc03} and \cite{ma05} SPS models can reproduce
the \bvm\ and \ubm\ colors in the selection box, the exact colors at
fixed age and star formation timescale are slightly different. In
order to test whether these small differences cause a bias that favors
a particular SPS model, we split the post-starburst galaxy sample at
\bvm$=0.55$, and repeated the analysis for both sub-samples. Both SEDs
are significantly better fit by the \cite{bc03} models (with similar
fit qualities as for the full composite spectrum), where the redder
sample is older by $\sim$0.1 dex. Additionally, we repeat the
analysis for a redder boundary \bvm$<0.6$, yielding a sample of
$\sim$110 galaxies. The composite spectrum is better fit by a
slightly older stellar population, but still strongly favors the
\cite{bc03} models above the \cite{ma05}. Altogether, this confirms
our unbiased selection.

\cite{mac10} compare the near-infrared fluxes of two spiral galaxies
to predictions from long-slit spectra, reporting better consistency
with the \cite{ma05} models than with \cite{bc03}. However, the
interpretation of their results is complicated by the presence of a
significantly older stellar population, which contributes 26\%--67\%
to the total light, and the presence of dust, which may mimic the
effects of a TP-AGB star population. Nonetheless, it is possible that
the \cite{ma05} model indeed provides a better description for the SED
shape during the evolution phase of these spiral galaxies.

\cite{ma06,ma07} use the broadband photometry of seven
spectroscopically confirmed galaxies to assess the fit quality of
different SPS models and find no overall preference for a particular
SPS model. Only one galaxy (ID~3650) of this sample would have
entered our selection (see Figure~\ref{fig:sel}). We have re-fitted this
galaxy using the same method as for the composite spectrum, and --
broadly consistent with \cite{ma06} -- we neither find a strong
preference for a particular SPS model (see Table~\ref{tab:mod}). Thus,
the differences in fitting techniques do not account for the large
discrepancy in $\chi^2$ value for the composite spectrum. Moreover, as
several galaxies in our sample, in particular at higher redshift, are
almost equally well fit by the \cite{ma05} models, our study is not in
disagreement with \cite{ma06}.

This redshift dependence may be caused by the fact that the
metallicity and dust content at fixed stellar mass evolve with
redshift \cite[e.g.,][]{er06,mai08}, and thus it may be possible that
the contribution from TP-AGB stars also changes with time. We test
this by splitting our sample at $z=1.56$ and repeating the analyses
for both sub-samples. Although, for both samples the \cite{bc03}
models provide significantly better fits, the relative fit quality of
the \cite{ma05} models versus \cite{bc03} increases with
redshift. Future work might be able to separate these trends and
constrain TP-AGB models as a function of stellar mass and metallicity.

\subsection{Utilizing a Flexible SPS model}

The difference in fit quality between the \cite{bc03} and \cite{ma05}
models for the composite post-starburst galaxy spectrum reflects the
different treatments of the TP-AGB phase. Our results suggest that the
treatment by \cite{bc03} is more appropriate for our post-starburst
galaxy sample than that of the \cite{ma05} models. We use the flexible
SPS\footnote{These models can be downloaded at
  www.cfa.harvard.edu/\~\ cconroy/FSPS.html.} (FSPS) models by
\cite{co09,co10} to obtain a more quantitative constraint on the
TP-AGB phase, as FSPS allows the modification of the bolometric
luminosity and effective temperature of the TP-AGB stars.

We start with the best-fit FSPS model to the composite post-starburst
galaxy spectrum, restricting the wavelength region to
$\lambda<6000$\,\AA\ and assuming the latest default Padova TP-AGB
models \citep{gi00,mg07,mar08}. Similar to the other models, this is a
stellar population with log\,($t$/yr)=9.04, log\,($\tau$/yr)=8.1, and
an \av\ of 0.4 mag (for solar metallicity). Next, we fix the age,
star formation timescale and dust content of the stellar population
and vary both the effective temperature and bolometric luminosity of
the TP-AGB stars, quantified as shifts with respect to the Padova
evolutionary tracks, \dt\ and \dl, respectively. Thus, we ignore any
potential degeneracies between \dt,
\dl, and other stellar population properties \citep{co09}. 

Figure~\ref{fig:fsps} illustrates the influence of both parameters on
the SED and shows their $\chi^2$ contours. The best fit has a reduced
$\chi^2$ value of 0.75, thus similar to \cite{bc03}. The
composite post-starburst galaxy spectrum suggests that the predicted
effective temperatures of TP-AGB stars in the Padova models are
consistent with our observations, but the overall luminosity is
$\sim$0.5 dex lower. 

\section{CONCLUSIONS} \label{sec:dis}

In this Letter, we define a photometrically selected sample of 62
post-starburst galaxies from the NMBS and use their composite SED to
obtain new constraints on the SED shape during the time that the
TP-AGB stars are thought to be most dominant. The SED is well fit by
the \cite{bc03} SPS models, while the \cite{ma05} models do not
reproduce the rest-frame optical and near-infrared parts of the SED
simultaneously, implying that these models give too much weight to
TP-AGB stars. This has previously been found by \cite{cg10} using
post-starburst galaxies in the SDSS. 

The high-resolution photometric sampling of the NMBS allows us to
derive quantitative constraints on the luminosity in the TP-AGB
phase. Using the FSPS models by \cite{co09}, we find that the
bolometric luminosity of TP-AGB stars is a factor of $\sim$3 lower
than predicted by the latest Padova TP-AGB models, when assuming solar
metallicity. However, an independent abundance measurement is needed
to break the degeneracy between metallicity and the TP-AGB phase
parameters. 

The significant reduction in the bolometric luminosity of TP-AGB stars
that is required to fit the observed post-starburst SED reflects one
or more failures of the current generation of SPS models. This
reduction can be physically achieved in the models by reducing TP-AGB
lifetimes, reducing bolometric luminosities of stars in the TP-AGB
phase, and/or embedding a significant fraction of TP-AGB stars within
optically thick circumstellar dust shells. These different
explanations may be disentangled by appealing to mid-IR data
\citep[see, e.g.,][]{kh10,sa09}.

Taking our results at face value, we infer that stellar masses and
other stellar population parameters derived using the \cite{ma05}
might be biased for certain evolutionary phases. We stress that our
findings do not imply that the \cite{bc03} models are best for all
galaxies or for all evolutionary phases. Using similar techniques as
presented in this Letter, in combination with flexible SPS models,
which allow unrestricted modification of uncertain evolution phases
and input parameters (such as metallicity), it will be possible to
substantially reduce many of the major uncertainties plaguing all SPS
models.

\acknowledgements We thank the referee for a very constructive report,
Jenny Greene, James Gunn, and Scott Trager for useful discussions, and
the COSMOS and AEGIS teams for the release of high-quality
multi-wavelength data sets to the community.

\end{document}